\documentclass{article}

\usepackage{arxiv}

\usepackage[utf8]{inputenc} 
\usepackage[T1]{fontenc}    
\usepackage{hyperref}       
\usepackage{url}            
\usepackage{booktabs}       
\usepackage{amsfonts}       
\usepackage{nicefrac}       
\usepackage{microtype}      
\usepackage{lipsum}		
\usepackage{graphicx}
\usepackage{natbib}
\usepackage{doi}

\title{The Footprint of Campaign Strategies in Farsi Twitter: A case for 2021 Iranian presidential election}

\author{\hspace{1mm}Saeedeh Mohammadi\\
Physics Department\\
Shahid Beheshti University\\
Tehran, Iran\\
Center for Complex Networks \& \\
Social Data Science, Tehran, Iran\\
\And
\hspace{1mm}Parham Moradi\\
Center for Complex Networks \& \\
Social Data Science, Tehran, Iran\\
\And
\hspace{1mm}S. Mahdi Firouzabadi\\
Institute for Cognitive \& Brain Sciences\\
Shahid Beheshti University\\
Tehran, Iran\\
\And
\hspace{1mm}G. Reza Jafari\\
\texttt{gjafari@gmail.com}\\
Physics Department \\
Shahid Beheshti University \\
Tehran, Iran \\
Irkutsk State Technical University\\
Lermontovastr. 83, 664074 Irkutsk, Russia\\}
\date{\today}

\renewcommand{\shorttitle}{\textit{arXiv}}

\begin{document}
\maketitle

\begin{abstract}
The rise of social media accompanied by the Covid-19 Pandemic has instigated a shift in paradigm in the presidential campaigns in Iran from the real world to social media. Unlike previous presidential elections, there was a decrease in physical events and advertisements for the candidates; in turn, the online presence of presidential candidates is significantly increased. Farsi Twitter played a specific role in this matter, as it became the platform for creating political content. In this study, we found traces of organizational activities in Farsi Twitter. Our investigations reveals that the discussion network of the 2021 election is heterogeneous and highly polarized. However, unlike other elections, candidates' supporters are very close, and "Anti-voters" who endorse boycotting the election is at the discussions opposite end. Furthermore, high presence of the bot activity is observed among the most influential users in all of the involved communities.
\end{abstract}

\keywords{Iran election \and Twitter \and complex networks \and polarization \and social manipulation \and echo chamber}

\section{Introduction}

The increasing number of online users who spend more and more time on social media has made these platforms a significant source of influence in modern societies.The transparency provided by social media created a sense of trust in users to use these platform as their source of information\citet{shirazi2013transparency}, therefore they have become a vital tool for political campaigns as a valuable source of information about voters' opinions and to deliver politically tailored messages. This paradigm shift instigated the emergence of organized influence operations on social media by politically motivated parties to promote their agenda \citet{mattei2021italian}. The notable example is the Russia-related influence operation to manipulate voters in the 2016 US presidential election,. There are  many other cases noted in numerous studies \citet{badawy2019characterizing,diresta2019tactics}. 
Employing fake users, trolls, and bots to propagate a political party's agenda has been used as a political tactic for different events \citet{gorwa2020unpacking}. Election campaigns use these methods, and sophisticated digital marketing techniques for their operations during elections \citet{bessi2016social,ferrara2017disinformation}. Detection of these organized activities and inauthentic users is significant in protecting the integrity of democracies. Various researches have employed network-based and content-based features to build novel computational models to identify online manipulations \citet{Grinberg2019, Badaway2018, alizadeh2020content}

Many studies have investigated the role of Twitter in Iranian diplomacy and public political debate. The Green Movement, which was a protest in response to the 2009 presidential election results, was praised by the Washington Times as "The Twitter revolution" \citet{washingtonpost}. At that time, Twitter facilitated coordinating the protests \citet{burns2009}. A year later, The Atlantic \citet{atlantic} published an article titled "Evaluating Iran's Twitter revolution" that criticized the premature claims previously made by the Washington Times. The article points out that social media was not responsible for mobilizing the protesters; however, it was vital in informing outsiders about the protest's events. Further studies support this assertion \citet{Howard2010,morozov2009}. Henceforward, according to some analysis, Farsi Twitter serves as a space for free expression of political opinions by the reformists or the opposition and as an alternative to the state's media \citet{khazraee2016,faris2015}. The articles published about the two previous presidential elections tried to eliminate bot activity in their analysis and focused on genuine users activity in Farsi Twitter\citet{khazraee2016,kermani2021mapping}. Both these studies mention that most of the content created in Farsi Twitter favors the reformist candidate. 

In this article, we examine the political debate in Farsi Twitter in the heat of the 2021 Iran presidential election. We use the trending hashtags regarding the election from seven-week before to one week after the election. While the opposition discourse was still active in the Twitter election debate, we observed a significant rise in tweets favoring the conservative candidates, particularly several days prior to the election. The retweet networks about this discussion is highly polarized, however the presidential candidates are together at one side and the other pole consists of the opposition.This could point to the political structure in the real Iranian political space\citet{saeedian2016visas}. 

We present evidence from network analysis suggesting high levels of user manipulation through hashtags led by inorganic users favoring almost all campaigns involved in the election. Our results imply that the trending hashtags of Farsi Twitter in such events merely represent public opinion, since users who tweet hashtags generally have a political agenda. Our study approves that analyzing network communities and their dynamics offers powerful means to investigate campaign strategies of different political groups.

\section{Methods}
In this section, we describe our approach in data gathering, then elaborate on the configuration of the networks, and finally explaining our method to analyze the networks based on network topology.

\subsection{Data Collection}

We monitored trending hashtags daily and gathered 97 hashtags related to the 2021 Iran presidential election from the 29th of April to the 24th of June (7 weeks before the election to one week after). These trending hashtags represented topics related to the election. Most Hashtags were either sign of support or opposition to a candidate or the whole election process. 

We constructed a "Twitter Machine" program that takes a keyword and an initial date as the input and stores any tweet, including that word, since the given date in an SQLite database. The tweets are made available through the Twitter's Standard Search API. Twitter Machine deconstructs the JSON that API returns records the necessary data in an SQLite database and updates the user index in the PostgreSQL database. The PostgreSQL is used as a relational database and contains every user participating in a conversation gathered by Twitter Machine. The users' information is regularly updated when Twitter Machine does a new task.  

The Twitter Machine collected 8818675 tweets that included the hashtags relating to the election discussion. 153115 users tweeted these posts over the 8 weeks used in this study.

In order to analyze the transformation of user behavior over time, these data was divided into eight different periods.Table \ref{tab:table1}, presents these 8 time frames. For each time frame, the amounts of
tweets and users that is collected by Twitter Machine are listed.

\begin{table}
\small\sf\centering
\caption{The different time frames of data collection are listed with the amounts of users and tweets collected during this time. BPE and APE stand for before and after presidential election respectively. \label{tab:table1}}
\begin{tabular}{clrc}
\toprule
\textbf{Label}&\textbf{Time Frame}&\textbf{No. users}&\textbf{No. tweets}\\
\midrule
a&7th-5th Weeks BPE&444,204&31,015\\
b&5th-4th Weeks BPE&893,868&47,316\\
c&4th-3th Weeks BPE&1,114,342&50,194\\
d&3th-2th Weeks BPE&1,591,757&65,325\\
e&2th-1th Weeks BPE&1,117,377&72,607\\
f&1th Week-2th Day BPE&1,722,812&74,464\\
g&2th Day BPE-1th Day APE&987,984&54,054\\
h&1th Day-1th Week APE&478,363&49,549\\
\bottomrule
\end{tabular}
\end{table}

To identify user authenticity, we used Botometer \citet{botometer} and gathered Botscore, Complete Automation Probability (Cap), and Fake Followers for users. To improve the accuracy of the analysis, we used crowdsourcing to annotate over 1000 accounts manually. As a result, we noticed that users with CAP scores less than 0.01 are likely to be genuine users, and the ones with CAP scores higher than 0.75 are more likely automated or semi-automated accounts.

\subsection{Networks}

\begin{figure}
\centering
  \includegraphics[width=0.8\linewidth]{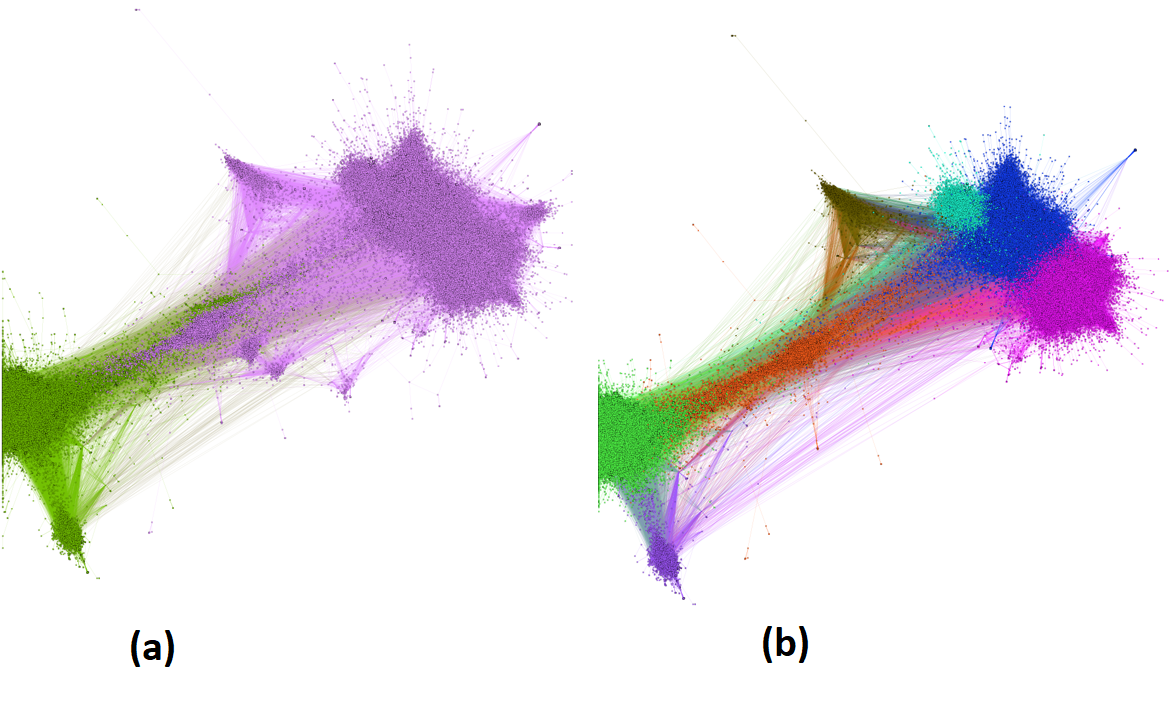}
  \caption{Retweet network of the election discussion in Farsi twitter with different resolutions when calculating the modularity. The colors represent the node's community that is detected by modularity. The size of the nodes is proportional to the accumulated number of retweets of the user. (a)resolution is set to 25 (b) resolution is set to 1. \label{figure:whole_network}}
\end{figure}

Network analysis can be applied to understand whether the election discussion and user behavior are well represented in the election Hashtags. While the friendship network embodies valuable information about the users' social interactions, the retweet network seems to be better for this study. It would reveal information about the discussions surrounding specific events of the election. To ensure that we would only include politically engaged users, each node in the retweet network is a user that tweeted a post that included a hashtag relevant to the election.
If user A retweets a post from user B at any point, an edge links user B to user A in the retweet network. In figure \ref{figure:whole_network}, the retweet network for the whole duration of data collection is illustrated using forceAtlas2 algorithm \citet{forceatlas2}, where the size of each node is proportional to the user's accumulated retweet count. Note that this number differs from node's in-degree since the network is constructed from only the tweets that include a trending hashtag about the election.

Node color in the network indicates to which community a node belongss. The community detection is done using modularity classes \citet{modularity}. 
A network with N nodes and L edges is divided into \(n_c\) subgraphs each of which \(L_c\) edges. The modularity of each subgraph is calculated by measuring the difference of the wiring of the existing network with a random network with the same amount of nodes and edges. Therefore, a cluster's modularity is defined as,
\[M_c=\frac{1}{2L} \sum_{(i,j) \in C_c} (A_{ij} - P_{ij}) \]
where c indicates subgraph, \(A_{ij}\) and \(P_{ij}\) are the elements of the adjacency matrix of the entire network and the random network, respectively. The cluster's modularity can also be calculated as follows;

\[M_c= \sum_{c=1}^{n_c} [\frac{L_c}{L} - (\frac{k_c}{2L})^2]  \],
where \(k_c\) is the sum of each node's degree within each cluster.

The resolution set in the modularity \citet{modularityres} specifies the size of the communities. The higher the resolution, the communities get more extensive and include more nodes. Eventually, The amount of communities (\(n_c\)) becomes smaller. Figure \ref{figure:whole_network}(a) illustrates the retweet network with a resolution set to 25 that has detected the two major communities, whereas, in figure \ref{figure:whole_network}(b) the communities are more but smaller with resolution set to 1.

Some of the network's essential structural characteristics are measured to study the structure of the retweet network over time. These properties include; Average clustering coefficient, modularity, average shortest path, and network assortativity.
Average clustering coefficient describes the closeness of each node's neighbours. Node i's clustering coefficient (\(c_i\)) is calculated by \(c_i = \frac{2L_i}{k_i(k_{i}-1)}\).
Where \(L_i\) is the number of edges between i's neighbors, and \(k_i\) is the node i's degree.

The average clustering coefficient of the whole network is the mean value of all the node's clustering coefficients\citet{clustering}.

Assortativity measures each node's likelihood to link with similar nodes to itself\citet{newman2002assortative}. A random network's assortativity is zero since the nodes are attached by random probability. Each node pair is assigned an assortativity coefficient,r, which is derived by the distribution of the remaining degrees,\(q_k\),and the joint probability distribution of the pair ,\(e_{jk}\)
\[r = \frac{\sum_{jk}[jk(e_{jk}-q_jq_k)]}{\sigma_q^2}\]
\[\sigma_q^2=\sum_k[k^2 q_k]-[\sum_k[k q_k]]^2\]

Many studies indicate a dual nature for Twitter. One as a medium for social communication, exchange of opinions with more reciprocal ties, and echo-chamber-like spaces. The other is a medium to receive information and as a news platform with more one-sided relations. To investigate this idea, we compare the reciprocal and non-reciprocal retweet networks of the election. The reciprocal network is a subgraph of the non-reciprocal network. Each edge in this network is among users whom at least retweeted each other once.

As Colleoni outlines it, the network based on reciprocal ties might indicate higher levels of homophily and echo-chamber-like environment\citet{colleoni2014echo}. We use the external-internal (E-I) index to measure the isolation and embedding of communities in each network. The I-E index is the difference between the number of ties with outside members and group members divided by the total number of ties. the total number of ties ranges from -1 (all ties are internal to the group) to +1 (all ties are external to the group) \citet{del2018echo}
As we expected, our results indicate a higher level of homophily in reciprocal networks which means that the two-way communication about the election is highly confined within same-minded isolated communities. 

In the following section, we will discuss the detailed properties of the two retweet networks.

\section{Discussion}

In this section, the retweet network of the election discussion is analyzed. A retweet network is considered to be the network of information dissemination. 

To detect the supporters of each candidate, first, we use modularity classes to separate nodes into different clusters (For this article, we only consider the seven largest communities and excluded smaller ones).
Next, the top 10 nodes with the most retweet from each community are manually investigated by going through the tweets they posted in that specific period. After detecting these user's political preferences, the clusters are labeled accordingly. 
All clusters labeled for a specific candidate or a political stance (Anti voting or Pro voting) are called their community. Afterward, Cocran's law is used in sampling the users in each community to measure the accuracy of community labels. The average accuracy for communities in all the retweet networks is above 80\%. In addition, as it gets closer to the election, the accuracy increases from 86\% to 93.5\%. This increment suggests the creation of echo chambers in the retweet network as we gets closer to the day of election.

\begin{figure}
\centering
  \includegraphics[width=0.7\linewidth]{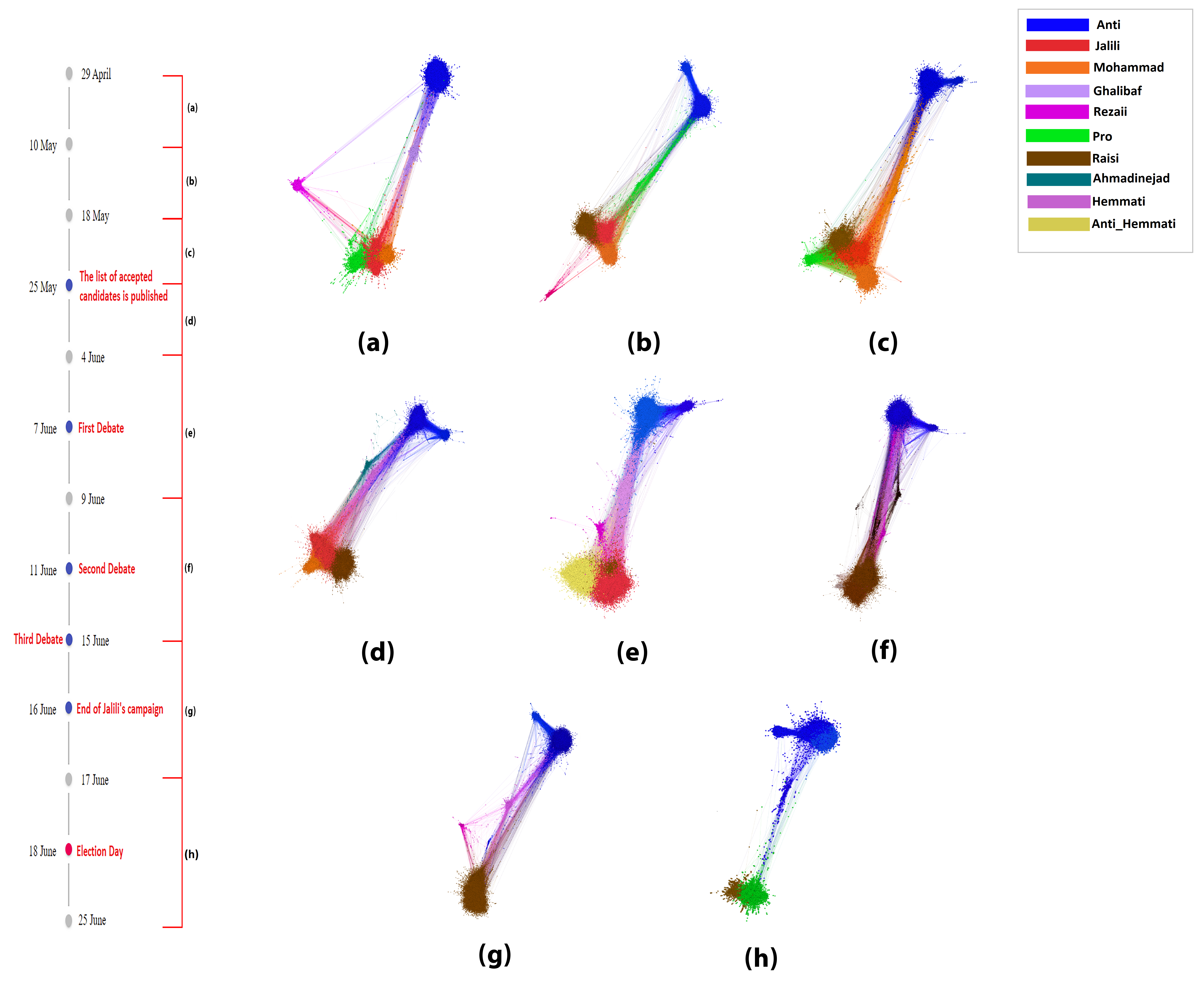}
  \caption{An illustration of the seven biggest communities of retweet networks throughout data collection. On the left, a timeline of dates and events surrounding the election is illustrated; The period of each retweet network is specified as well. Each node's color corresponds to the community to which it belongs. \label{fig:rtntrw}}
\end{figure}

In figure \ref{fig:rtntrw}, the retweet networks are illustrated with ForceAtlas2 \citet{forceatlas2} as the layout algorithm. Each community is defined with different colors. Each network represents the discussion about the election in a particular period, specified in the timeline on the left.

The anti-community (users that declare their refusal to vote and encourage others to boycott the election) is on one end of the discussion and the opposite end, the pro community (users that encourage others to vote and claim they would as well), Jalili community, Mohammad community, and Raisi community. This state is valid throughout the timeline. While Mohammad and Jalili communities are not active in the days closer to the election and after, the Raisi community preserves their place opposite the anti-community.

It is evident in figure \ref{fig:rtntrw} that the anti-community constitutes multiple clusters in most networks. The community detection shows 2 clusters in the anti-community, which we call; the main body and the tail. The hashtags widly used in both of these clusters are examined. The most used hashtags in the tail cluster are hashtags that are more likely used by MEK supporters, which would suggest that this cluster belongs to MEK supporters. However, since both these clusters propagate the same message, boycotting the election, we refer to both as the anti-community.

After the process of the candidate validation was ended, the approved candidates' list was published on the 25th of May, many supporters were disappointed as their candidate could not participate in the election. Therefore, they expressed their frustration on Twitter. The Ahmadinejad community presence in the retweet network (d), close to the anti-community and between the poles, points to the anger of 
his supporters over the decision. On the other hand, Mohammad's supporters, who have also declined a presidential candidacy, preserve their place opposite the anti-community and close to Jalili and Raisi's communities.   

\begin{figure}
\centering
  \includegraphics[width=0.5\textwidth]{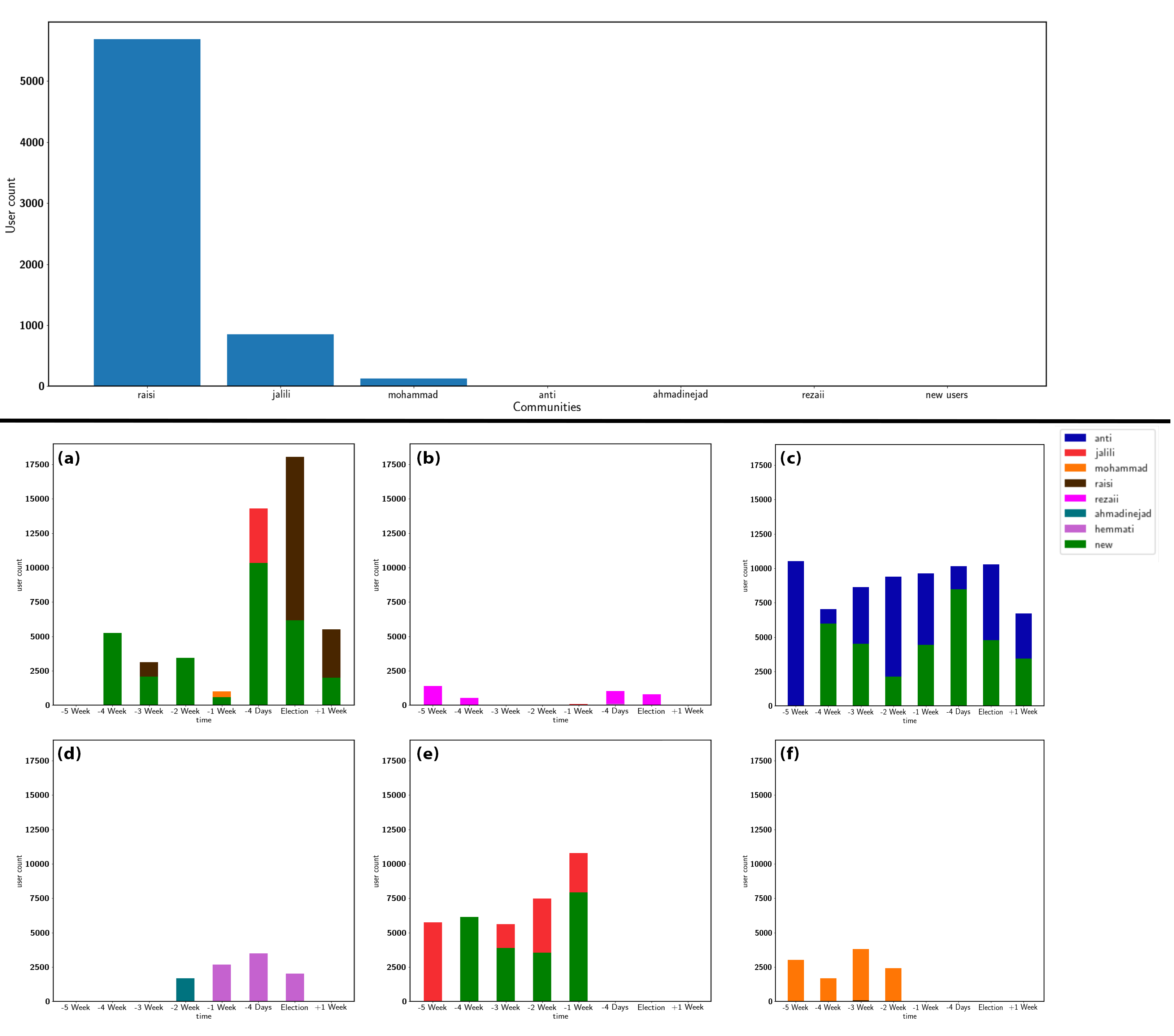}
  \caption{(Above) User distribution of the anti-hemmati community. (Below) The user distribution over the timeline in each community. The colors refer to which community the users belonged to in the previous tick.The communities are: (a)Raisi, (b)Rezaii, (c)Anti, (d)Hemmati (e)Jalili (f)Mohammad. \label{fig:political_affiliation}}
  
\end{figure}

During the 2021 presidential election debates, many hashtags started trending in opposition to the Reformists candidate, Hemmati. The content of these hashtags did not show any support for any particular candidate and just criticized Hemmati. Hence, users who used these hashtags during the debates are labeled "anti-hemmati," their community is observed in the (d) network. It is important to note that this community is very close to Jalili and Raisi's community. Figure \ref{fig:political_affiliation}(Above), demonstrates which community these users belonged to in the previous network, which shows most of them belonged to Raisi community. This could explain the reduction of Raisi's community in this network. For future investigations, we would merge "anti-hemmati" and Raisi's communities.

\begin{figure}
\centering
  \includegraphics[width=0.5\textwidth]{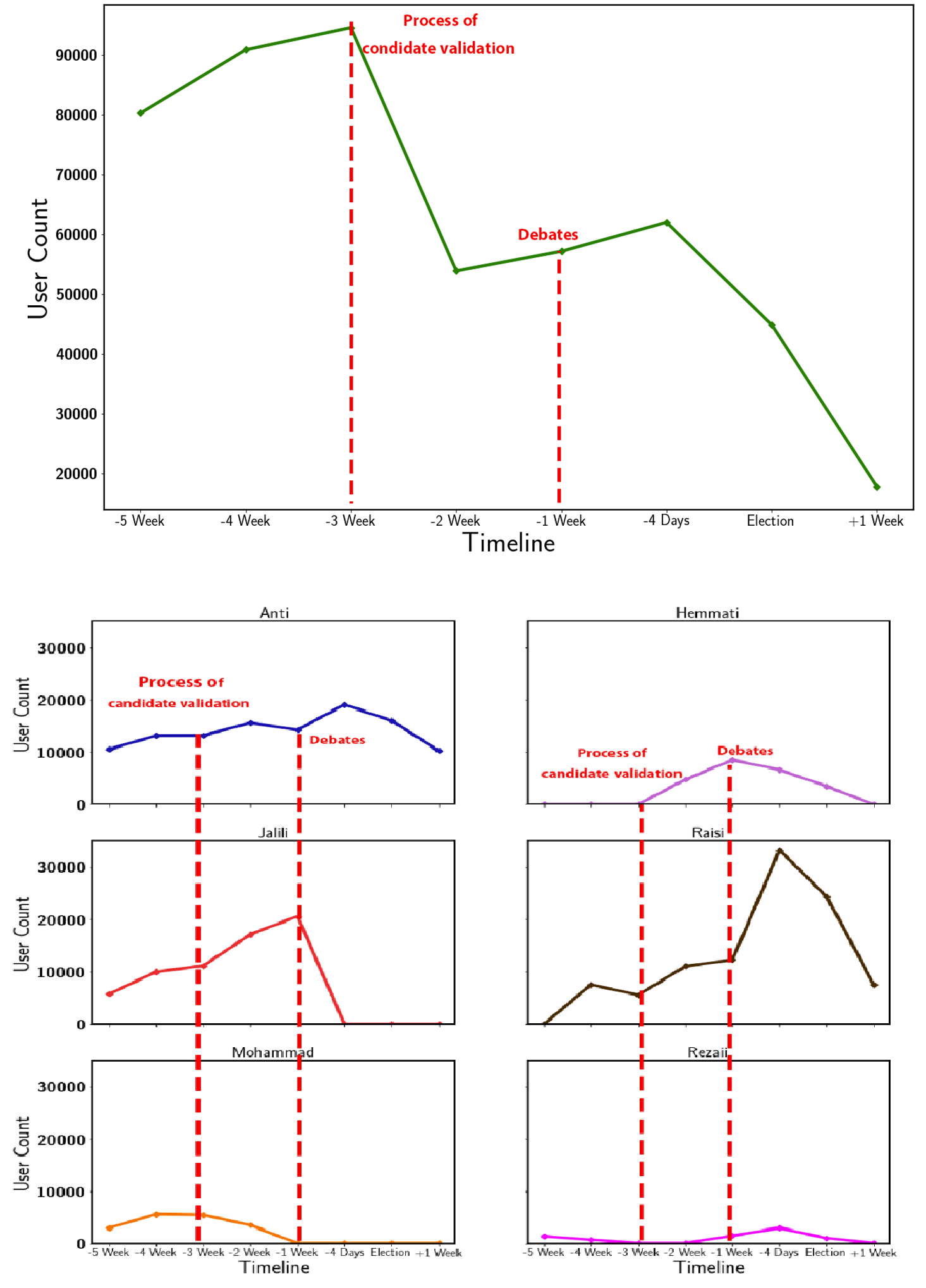}
  \caption{Top: The amount of users that participated in the election discussion by tweeting a trending hashtag. Bottom: The amount of participating users separated by communities. \label{fig:user_count}}
\end{figure}

As it gets closer to the election itself, the supports for some candidates are reduced and increased for some others. For instance, Mohammad’s community is not in the retweet network after the (d)
network or Rezaii’s community disappears during (c) and (d) network and reappears again after that. It is important to note that the fact that these communities are not in the retweet networks in figure\ref{fig:rtntrw}, does not mean that they were no support for them in Twitter during that time, it just implies that the support for them was reduced to the point that they did not appear in the seven biggest modularity class of the original retweet network. figure \ref{fig:user_count}, offers the amount of users involved in the election discussion in Farsi Twitter during the data collection period. The user count of each community has been done for each time period.

An interesting finding from figures \ref{fig:rtntrw} and \ref{fig:user_count} is in the (f) network, one week to 4 days before the election, where Jalili's supporters decrease significantly and are omitted from major communities. This omission happens while Jalili ends his campaign two days later, which could suggest that this sudden lack of support in social media compels Jalili to end his campaign, or the decision was made a week prior to the election systematical move from Jalili's team. Furthermore, there is a significant rise in Raisi's community as Jalili's decline, suggesting that the users who formerly belonged in Jalili's community have merged with Raisi's supporters. This claim is proved in figure \ref{fig:political_affiliation} (Below).

Figure \ref{fig:political_affiliation} depicts the way communities have grown or contract during this time frame. Each bar shows the number of users in a specific community, and the colors illustrate which community they belonged in the previous tick. New users who were not in the previous communities are also demonstrated. In figure \ref{fig:political_affiliation} (a), it is shown how Jalili's community merged into Raisi's community four days before the election; the same also happens as Mohammad's community merges into Raisi's community when Mohammad's community declines. In addition, the lack of new users or users from other communities in Rezaii and Mohammad's community could point to an organized activity.

\begin{figure}
\centering
  \includegraphics[width=0.4\textwidth]{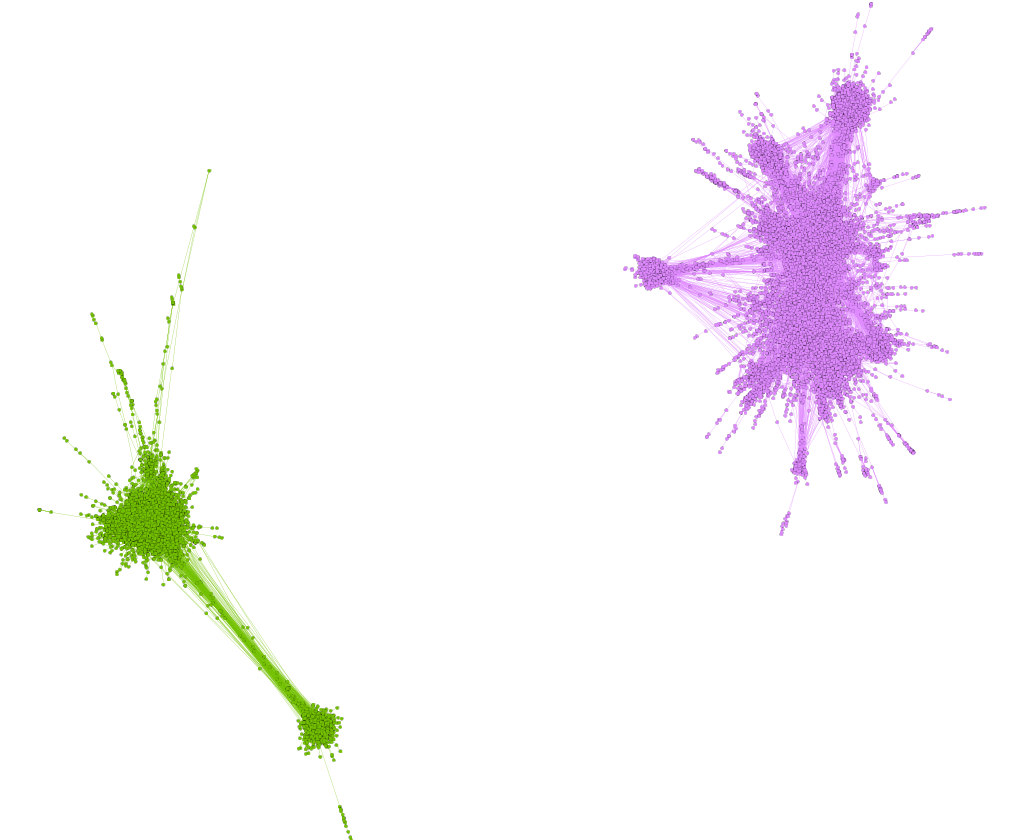}
  \caption{The reciprocal subgraph of the retweet network of the election discussion for the whole time of data gathering \label{fig:reciprocal_network}}

\end{figure}

\begin{figure}
\centering
  \includegraphics[width=0.8\linewidth]{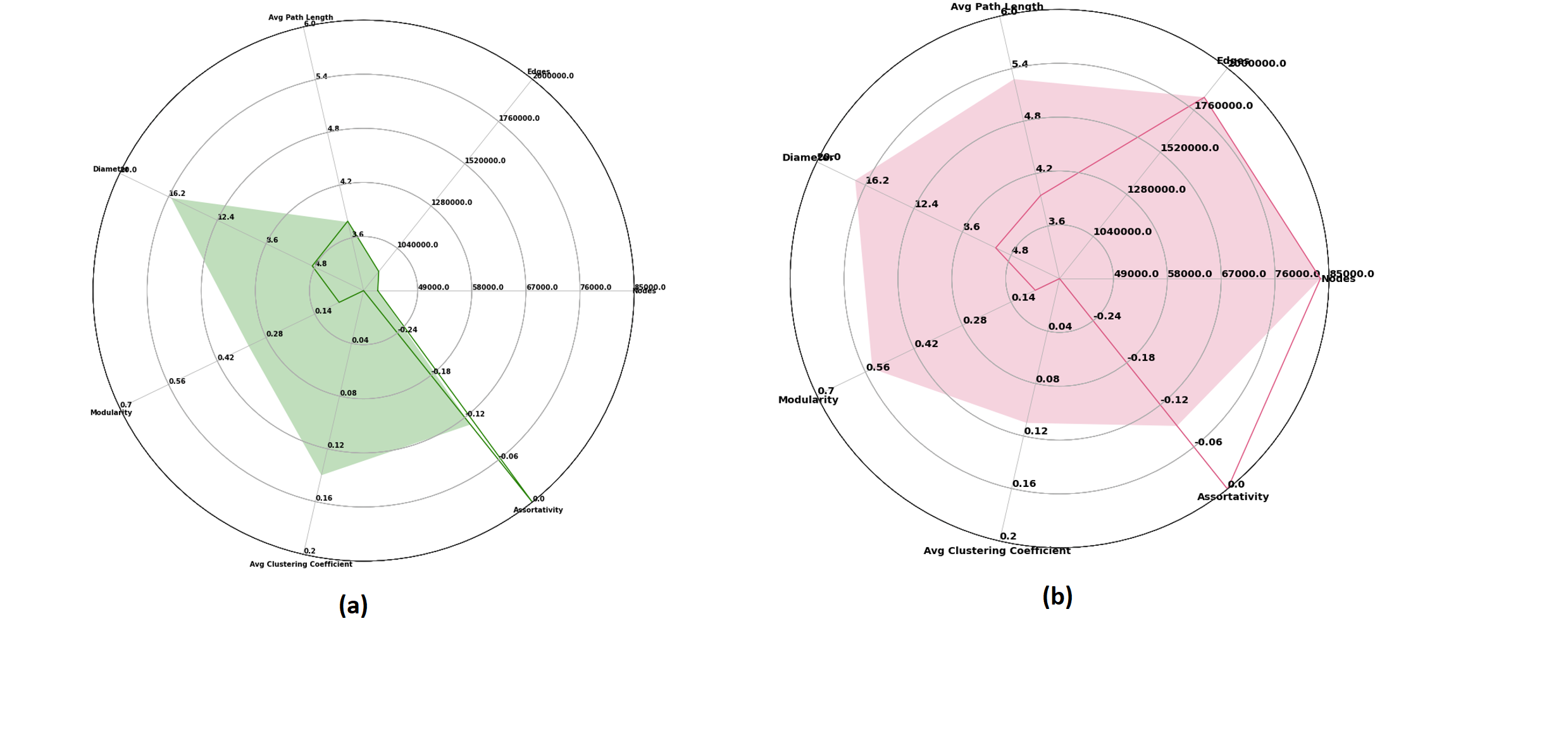}
  \caption{radar graph of structural properties the two main subgraphs in the election discussion. The properties are; number of nodes, number of edges, diameter, average path length, modularity, assortativity, average clustering coefficient. The solid line represents the properties of a random graph with the same amount of nodes and edges. (a) Anti subgraph (b) Pro subgraph.  \label{fig:radar}}
  
\end{figure}

Figure\ref{figure:whole_network} (a), illustrates two central communities of the whole network. The I-E index of this network is about 0.98, which indicates the network is very polarized and heterogeneous. Figure \ref{fig:reciprocal_network} illustrates the reciprocal subgraph of this network, which shows that reciprocal ties are only within the two poles, and there is no such activity in between the poles. The communities of this network are labeled using the method mentioned before. The accuracy of the community labelling measured in this network is 100\%, which suggests that the dark pink community in the graph are users supporting a candidate and are "pro voting," the green community is users who are against the election and are "anti voting."
Although the anti subgraph is much smaller than the Pro subgraph, the structure of the poles is very similar. Figure \ref{fig:radar}, shows the radar graph for the pro and anti-community. It is demonstrated that in both these subgraphs; First, the assortativity coefficient is negative, suggesting that hubs are more likely to link to low degree nodes. Second, properties such as; diameter, average clustering coefficient, and modularity all deviate significantly from a random graph with the same number of nodes and edges. Lastly, the modularity of the Pro subgraph is much more than the modularity of the anti subgraph, which indicates that there are more evident communities in the Pro subgraph.

In order to identify the type of users who were significantly influential in the election discussion, Botometer API \citet{botometer} is used. In this article, the top 1 percent of most retweeted users in each community in each period is regarded as the most influential users. The Cap score and the relative score for fake followers are extracted from Botometer API for all the most influential users. Cap score is a Complete Automated Probability of boxscore that indicates what percentage of users with higher scores are bots. The score for fake followers is a relative number between 0 to 5, which specifies the number of fake followers each user has.

\begin{figure}
\centering
  \includegraphics[width=0.8\linewidth]{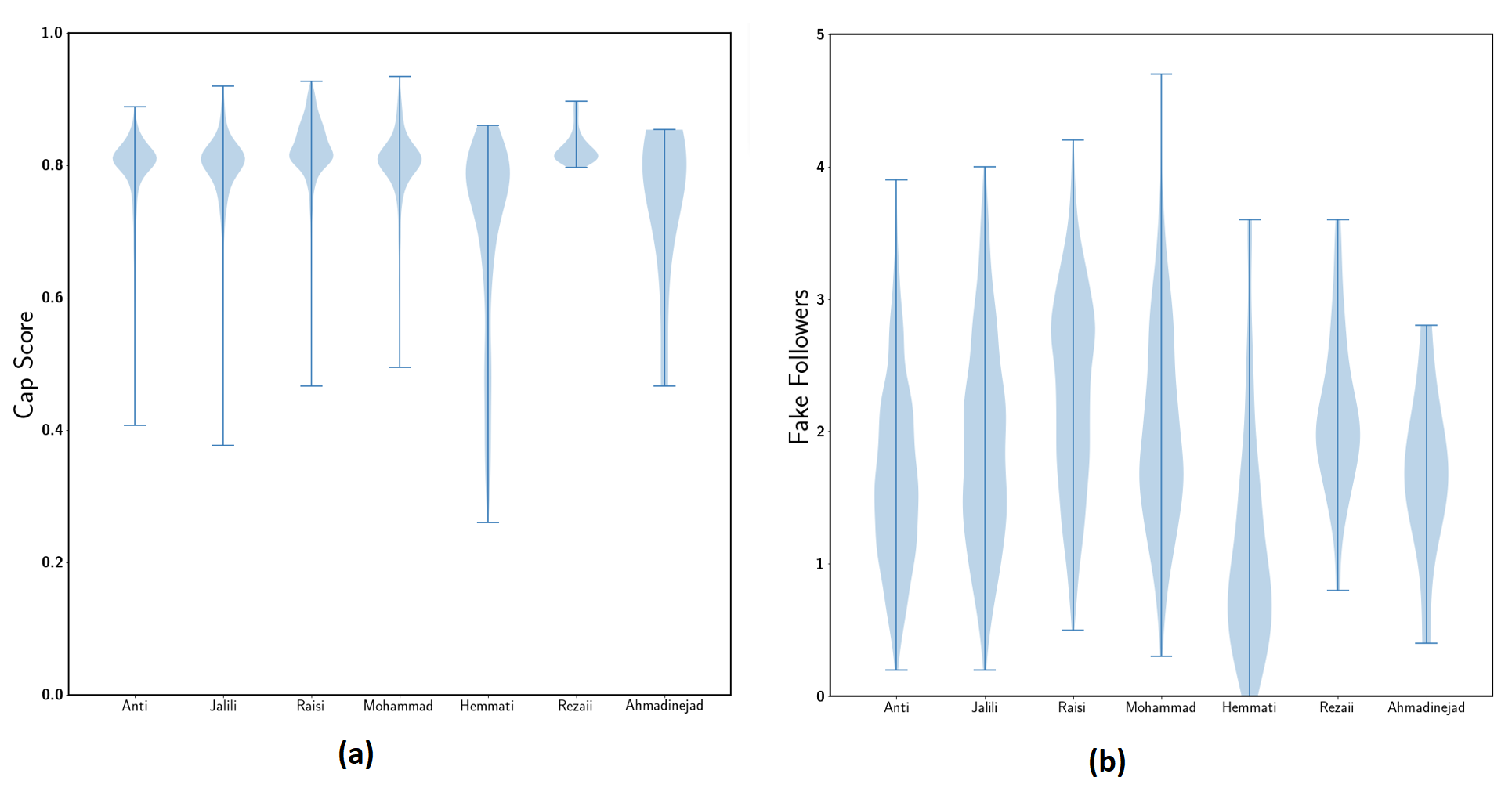}
  \caption{(a) CAP (Complete Automation Probability) Score for the top 1 percent most tweeted users from each community (b) The relative score for fake followers for the top 1 percent most retweeted from each community. \label{fig:total_cap}}
\end{figure}

Figure \ref{fig:total_cap} consists of violin plots that demonstrate the numerical distribution of Cap score and fake followers of the most influential users of each community. The significantly high Cap score among all the communities stipulates the remarkable presence of bots among these users.

The presence of bots is highly significant among communities involved in both ends of the network, i.e., Anti and Raisi, Jalili, and Mohammad. However, two of the communities in between the poles, Hemmati and Ahmadinejad's, have notably lower cap scores that point to the presence of more genuine users among their outspoken supporters.Their fake followers' relative score supports this claim is also noticeably lower than other communities. On the other hand, for Rezaii's community that was also mainly located between the two ends of the network, Figure\ref{fig:total_cap} shows a concentrated distribution of high cap scores in the most influential users that can be construed as most of Rezaii's influential supporters are either campaign accounts or bots. The reason behind the high botscore of users with the most retweets might be because this type of user acquires other non-genuine users to retweet them to propagate their message \citet{luceri2019red,botometer}; however, the mere presence of these accounts in some communities more than others points to organized activity to influence public opinion by these political groups.

\section{Conclusion}

As the world evolves into being digitized, the impact of social media platforms in socio-political is getting more vital than ever before. The covid-19 pandemic has accelerated this transition to online platforms. In the Iran 2021 presidential election, there was a noticeable rise in political content, specifically Farsi twitter, whereas the physical events and advertisements were significantly reduced. While this paradigm shift has provided necessary space for free discussion in a pandemic era, it has also converted a lot of crucial elements of the physical world into different things. For instance, in previous elections, endorsement from established political figures had a remarkable influence on voting behavior. However, social media has provided a space for anonymous users to propagate their ideology and influence elections. Considering the ban on Twitter in Iran, anonymity may not be an ill-nurtured act, more so an act of preserving user’s safety.

The results indicate a high presence of bot activity among the most outspoken users of all communities during the 2021 election, suggesting organized activity. The retweet networks of the elections' discussions preserve the same structure over the period leading to the election: a polarized network with one pole representing the supporters of presidential candidates and the other pole for users promoting the boycott of the election. This arrangement is an irregular observation comparing to other presidential elections, where the polarization is among candidates from opposing political parties. Even though the retweet network was not deformed as we got closer to the election day, we observed users' migration from different communities to the president-elect. At one point, the community that represented supporters of the presidential candidate, Saeed Jalili, merged into supporters of Ebrahim Raisi, president-elect, four days before Jalili officially suspended his election campaign. This could indicate that he opted out of the election due to his supporters' migration, or it could be a tactical move from his campaign. To summarize;

\begin{itemize}
 \item The paradigm shift from traditional media to social media opens up space for anonymous celebrities to influence public opinion. This anonymity, specifically in Farsi Twitter, is not considered to indicate bad intentions since Twitter is banned in Iran. Therefore, these new celebrities could be genuine users or bots. In the 2021 Iranian presidential election, it is concluded that the majority of the most outspoken users in the election discussion were not genuine users. 
 \item The anonymity provided by social media gives the opposition a platform to propagate their agenda.
 \item There are pieces of evidence that Iranian politicians in all groups have noticed the influence of social media in public opinion and have invested in infrastructure for an online movement in their favor.
 \item The rivalry of the 2021 election was not between the presidential candidates but between the group encouraging people to vote and the other that encouraged them not to. In other words, this election was a battle between the opposition and most of the presidential candidates.
 \item It is evident that all the invested parties have used campaign accounts or bots to propagate their agenda in this election. Since the most retweeted accounts in all communities, these accounts have an average score of 0.85, above the threshold for non-genuine users.
 \item The retweet network of the election discussion in Farsi Twitter is highly polarized with an I-E index of 0.98, and the reciprocal ties are only present within the poles and in the communities between the poles.
\end{itemize}

\section{Acknowledgments}
We thank Prof. Shant Shahbazian, Prof. Afshin Montakhab and Prof. Behzad Ghanbarian for their guidance and constructive comments regarding this manuscript.

\bibliographystyle{unsrtnat}
\bibliography{references}  
\section{Figures}


\end{document}